\documentclass[aps,pra,twocolumn]{revtex4-2}
\usepackage{bm}
\usepackage{amsmath}
\usepackage{amsfonts}
\usepackage{amssymb}
\usepackage{graphicx}
\usepackage{color}
\usepackage{xcolor}
\usepackage[colorlinks,linkcolor=blue,anchorcolor=blue,citecolor=blue,urlcolor=blue]{hyperref}
\usepackage{dcolumn}
\usepackage{fancyhdr}
\usepackage[normalem]{ulem}

\begin{document}

\title{Spin-orbit-coupled spinor gap solitons in Bose-Einstein condensates}

\author{Jing Yang}
\affiliation{Department of Physics, Shanghai University, Shanghai 200444, China}

\author{Yongping Zhang}
\email{yongping11@t.shu.edu.cn}
\affiliation{Department of Physics, Shanghai University, Shanghai 200444, China}

\begin{abstract}
		
Spin-1 spin-orbit-coupled spinor Bose-Einstein condensates have been realized in  experiment. We study spin-orbit-coupled spinor gap solitons in this experimentally realizable system with an optical lattice. The spin-dependent parity symmetry of the spin-orbit coupling plays an important role in the properties of gap solitons. Two families of solitons with opposite spin-dependent parity are found. Using an approximate model by replacing the optical lattice with a Harmonic trap, we demonstrate the physical origin of the two families. For the zero effective quadratic Zeeman shift, we also find a type of gap soliton that spontaneously breaks the spin-dependent parity symmetry.


\end{abstract}
	
\maketitle

\section{introduction}
	
In nonlinear phenomena, solitons are notably interesting~\cite{Kivshar1989, Haus1996}. Nonlinearities are able to balance the dispersion of spatially localized wave packets to form solitary waves. Depending on the physical origin and nonlinear backgrounds, the types of solitons are very diverse~\cite{Kartashov2011, Konotop2016}.  Among them, there is a particular family, called gap solitons~\cite{Chen1987, Mills1987}. The existence of gap solitons requires that there must be energy gaps in the dispersion relation of corresponding linear systems. Nonlinearities excite gap solitons and exactly situate them into these linear energy gaps.  In general, periodic potentials provide an important approach to open gaps in a linear dispersion relation. Therefore, nonlinear periodic systems are widely used to explore gap solitons. In nonlinear optics, periodic potentials can be afforded by waveguide arrays and optically induced photonic lattices~\cite{Lederer2008}. In atomic Bose-Einstein condensates (BECs), optical lattices are a powerful experimental means to control dispersion relation for the gap opening~\cite{Morsch2006}.

Besides their specific location in the linear gaps, gap solitons have a generic feature: they can be classified as fundamental and higher-order modes~\cite{Zhang2009PRA}. Higher-order modes can be considered as the composites of fundamental ones~\cite{Zhang2009,Alexander2011, Bersch2012}. Different from usual bright solitons that are supported by only attractive nonlinearities, gap solitons can exist in the presence of attractive~\cite{Mandelik2004,Neshev2004} or repulsive~\cite{Smirnov2007} nonlinearities. This makes it possible to experimentally observe gap solitons in  atomic BECs with repulsive interactions~\cite{Eiermann2004}.

The experimentally tunable optical lattices and interactions make  the BECs  an ideal platform to investigate gap solitons~\cite{ Louis2003, Efremidis2003, Lee2003, Ahufinger2005, Wang2008}. Furthermore,  one of the outstanding properties of the BEC systems is the experimentally possible implementation of multiple components.
Interactions and couplings~\cite{Mardonov2019} between multiple components bring novel properties into gap solitons~\cite{Ostrovskaya2004, 1Gubeskys2006, 2Gubeskys2006,Kivshar2007, Adhikari2009, Chen2017}.  Recently,  gap solitons in spin-orbit-coupled two-component BECs loaded into optical lattices have demonstrated interesting features originating from spin-flip symmetries of the spin-orbit coupling~\cite{Kartashov2013, Lobanov2014, Zhang2015,Sherman2019}. The study of the spin-orbit-coupled gap solitons is encouraged by the great achievement and  rapid development of experiments on spin-orbit-coupled BECs~\cite{Spielman2011, Goldman2014, JingZhang, Zhai, Li2015,  Shizhong2015, Yongping2016}. The research interest along the direction of spin-orbit-coupled gap solitons gradually increases~\cite{Chen2014,1Zhu2017, yongyaoli2017,Xu2018,  Sakaguchi2018, Kartashov2019, Fan2020, Zezyulin2020,Su2021}.

In this paper, we study the existence and properties of gap solitons in a spin-1 spin-orbit-coupled spinor BEC in the presence of optical lattices. The study is stimulated by the following aspects.
First,  a two-component spin-orbit-coupled BEC has been successfully loaded into optical lattices in the experiment~\cite{Hamner2015}.  Meanwhile, the spin-orbit coupling has been experimentally synthesized into a spin-1 spinor BEC~\cite{Campbell2016}. These experimental advancements inspire people to study the physics of a spin-orbit-coupled spinor BEC with optical lattices. Second, in comparison with two-component BECs, the spin-1 spinor BECs have more degrees of freedom,  and spinor interactions include the spin-spin collision which allows the spin exchange~\cite{Kawaguchi2012}. Many significant spin-mixing solitonic states have been experimentally observed in spinor BECs~\cite{Bersano2018, Chai2020}. Spinor gap solitons have already shown interesting spin structures~\cite{Kivshar2007}.  The interplay between the symmetries of the spin-orbit coupling and  spinor interactions is expected to give rise to more distinct spin states.  Finally, the spin-orbit-coupled optical lattices can have an alternative interpretation by the aid of the synthetic dimension concept~\cite{Boada2012,Celi2014}. Based on this idea, spin states become discrete sites along synthetic dimensions. The coupling between them turns out into the tunneling of the sites. The connection of the spin-orbit coupled optical lattices and the synthetic dimension has been experimentally examined~\cite{Livi2016}. Using the synthetic dimension, the system of spin-1 spin-orbit-coupled spinor optical lattices can be mapped into a model of three-leg ladders with a tunable magnetic flux~\cite{Zhehan2021,Mancini2015, Stuhl2015}. Such mapping enables experimental accessibility to chiral edge states by the standard spin-resolved atomic measurements~\cite{Mancini2015, Stuhl2015}. The study on spatially localized states in the spin-1 spin-orbit-coupled spinor optical lattices may shed light on the existence of novel quantum many-body edge states.

Our study completely bases on the spin-1 spin-orbit-coupled spinor BEC experiment of  $^{87} \text{Rb}$ atoms in Ref.~\cite{Campbell2016}. The spin-orbit coupling is synthesized by Raman lasers. It respects a spin-dependent parity symmetry.  We find that this symmetry plays an important role in the existence of the spin-orbit-coupled spinor gap solitons and brings interesting features into them. In the experiment of the spin-1 spin-orbit-coupled spinor BECs, the tunable parameters are the effective quadratic Zeeman shift and the Raman coupling. We find that the different effective quadratic Zeeman shift leads to diverse solitons.  For the zero effective quadratic Zeeman shift,  the population in the $|m_\mathrm{F}=0\rangle$ component is not preferred, and  gap solitons have two different types;
one type occupies the $|m_\mathrm{F}=-1\rangle$  and $|m_\mathrm{F}=1\rangle$  components and obeys the spin-dependent parity symmetry, while the other just occupies the one of  $|m_\mathrm{F}=-1\rangle$  and $|m_\mathrm{F}=1\rangle$ components and spontaneously breaks the spin-dependent parity symmetry. For a large effective quadratic Zeeman shift, we uncover two different families of gap solitons having opposite spin-dependent parity symmetries.  For a further understanding, we develop an approximating model to provide a physical picture for the origin of these two families.  All solitonic solutions can exist in a broad range of the Raman coupling and are stable. Their features are identified.

This paper is organized as follows. In Sec.~\ref{Model},
we present the theoretical model for the spin-1 spin-orbit-coupled spinor BEC with optical lattices. In Sec.~\ref{Linear}, the linear properties of the spin-orbit-coupled spinor optical lattice are studies, with a particular attention on the identification of linear energy gaps which shall accommodate gap solitons. In Sec.~\ref{NegativeZeeman}, two different types of spin-orbit-coupled spinor gap solitons are demonstrated in the case of the zero effective quadratic Zeeman shift. In Sec.~\ref{ZeroZeeman}, we demonstrate the existence of two families of solitons which have opposite spin-dependent parity symmetries for the case of a large effective quadratic Zeeman shift. An approximating model is developed to explain the origin of solitons.
Finally, the conclusion including a discussion of  ``edge state soliton"  follows
in Sec.~\ref{Conclusion}.

\begin{figure}[!t]
	\includegraphics[width=3.2in]{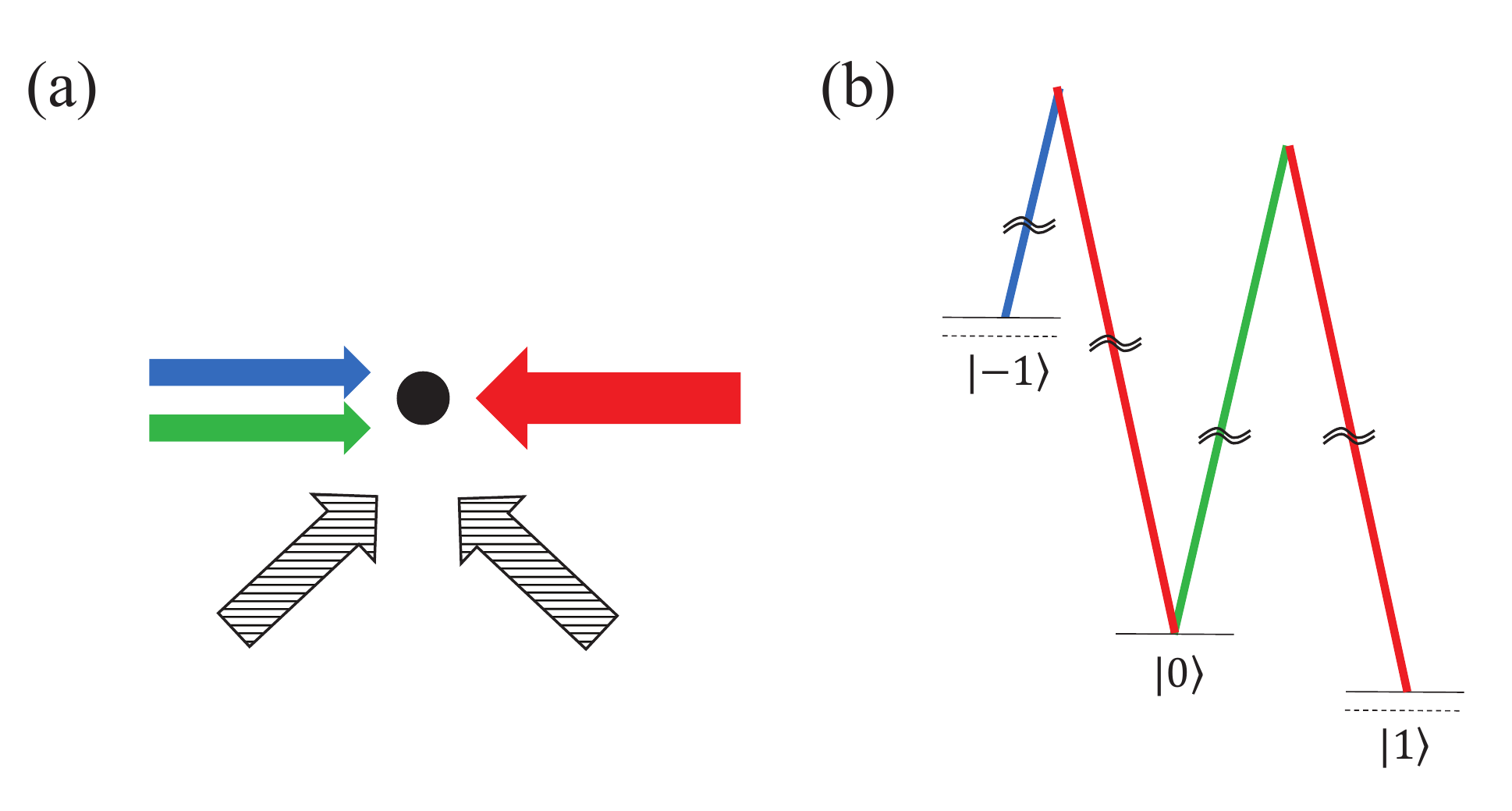}
	\caption{ Scheme for the realization of the spin-1 spin-orbit-coupled spinor optical lattice. (a) The spin-orbit coupling is implemented by the interactions between the atom cloud and three Raman lasers represented by the arrows in the horizontal direction, the two of them propagate in the same direction, and the third one in the opposite direction. Two additional far detuning lasers with a $\pi/2$ incident angle represented by the black arrows generate the optical lattice. (b) The interactions of  three Raman lasers and the unequal splitting hyperfine states of $^{87} \text{Rb}$ atoms. Two sets of two-photon transitions share the same laser represented by the red lines.  }
	\label{Fig1}
\end{figure}

\section{Model}
\label{Model}

Our system is quasi-one-dimensional. The spin-orbit coupling and the optical lattice are along the longitudinal direction, and the motions along the transverse direction are completely frozen due to strong traps. Ref.~\cite{Campbell2016} has experimentally realized a spin-1 spin-orbit-coupled spinor BEC. We follow the experimental design. The scheme to realize the spin-1 spin-orbit-coupled spinor optical lattice is shown in Fig.~\ref{Fig1}. With a large bias magnetic field, three hyperfine states of  $^{87} \text{Rb}$ atoms ($|-1\rangle=|\mathrm{F}=1,m_\mathrm{F}=-1\rangle$, $|0\rangle=|\mathrm{F}=1,m_\mathrm{F}=0\rangle$, and $| 1
\rangle=|\mathrm{F}=1,m_\mathrm{F}=1\rangle$) are split unequally between $|-1\rangle \leftrightarrow |0\rangle$ and $|0\rangle \leftrightarrow |1\rangle$. Three external Raman lasers (with the wavelength $\lambda_\text{Ram} =790 \text{nm}$) are employed to interact with the atom cloud. Laser frequencies and polarizations are properly chosen such that the Raman lasers can couple the hyperfine states together via two sets of two-phonon transitions [see Fig.~\ref{Fig1}(b)].  Two of the lasers propagate in same direction and the third one propagates oppositely. Such arrangement of laser propagating configuration introduces a spin-momentum locking during the transitions. The resulted single-particle spin-orbit-coupled Hamiltonian is \cite{Campbell2016},
\begin{equation}
	H_\mathrm{soc}=\frac{1}{2m}\left( p_x +2\hbar k_\mathrm{Ram} F_z \right)^2
	+\delta F_z^2+\sqrt{2} \bar{\Omega} F_x.
	\label{socH}
\end{equation}
In above, $p_x$ is the momentum along the longitudinal direction, and $m$ is atom mass. $F_z$ and $F_x$ are the spin-1 Pauli matrices. $\left( p_x +2\hbar k_\mathrm{Ram} F_z \right)^2/2m=p_x^2/2m +2\hbar k_\mathrm{Ram} p_xF_z/m +2\hbar^2  k_\mathrm{Ram}^2F_z^2/m $.  The spin-1 spin-orbit coupling is $2\hbar k_\mathrm{Ram} p_xF_z/m$ with the strength $2\hbar k_\mathrm{Ram}/m$, here $k_\mathrm{Ram}=2\pi/\lambda_\mathrm{Ram}$. $\delta$ relates to the two-photon detuning. In the experiment~\cite{Campbell2016}, the detunings for the two sets of two-photon transitions are adjusted to be equal, which results in $\delta F_z^2$. Incorporating  $2\hbar^2  k_\mathrm{Ram}^2F_z^2/m $ from the first term in $H_\mathrm{soc}$, the effective quadratic Zeeman shift becomes  $\left( 2\hbar^2  k_\mathrm{Ram}^2/m+\delta \right) F_z^2  $.  The last term in $H_\mathrm{soc}$ is the so-called Raman coupling with the strength $\bar{\Omega}$. 

Like the experiment of the two-component spin-orbit-coupled optical lattices in Ref.~\cite{Hamner2015}, we load such spin-1 spin-orbit-coupled system into an optical lattice by shining two $\lambda_\mathrm{lat}=1540\text{nm}$ lattice lasers. There is a $\pi/2$ incident angle between these two lasers which are represented by the black arrows in Fig.~\ref{Fig1}. The generated optical lattice is spin-independent and becomes $V(x)=-\bar{V}\cos(2k_\mathrm{lat} x)/2$ with $k_\mathrm{lat}=2\pi/(\sqrt{2} \lambda_\mathrm{lat}  )$. $\bar{V}$ is the lattice depth which can be tuned by changing the intensity of the lattice lasers.  Consequently, we end up with the total single-particle Hamiltonian for the spin-orbit-coupled spinor optical lattice,
\begin{equation}
H_\mathrm{sin}=H_\mathrm{soc}-\frac{\bar{V}}{2}\cos(2k_\mathrm{lat}x).
\label{singleparticle}
\end{equation}

The spin-1 BEC with the spin-orbit-coupled optical lattice is described by the standard mean-field Gross-Pitaevskii (GP) equations with the above single-particle Hamiltonian. Since the system has three components, the GP equations are three coupled nonlinear equations as follows,
\begin{equation}
\begin{aligned}
& i\frac{\partial\psi_{-1}}{\partial t}= H_{-} \psi_{-1}+\Omega \psi_0 
 +\eta \psi_0^2 \psi_{1}^*,  \\
& i\frac{\partial\psi_{0}}{\partial t} 
=H_0 \psi_{0}+\Omega (\psi_{-1}+\psi_{1}) 
	+2\eta \psi_{-1} \psi_{0}^*\psi_{+1}, \\
	& i\frac{\partial\psi_{1}}{\partial t}= H_{+} \psi_{1}+\Omega \psi_0 
	+\eta  \psi_{-1}^*\psi_0^2, 
	\label{GP}
\end{aligned}
\end{equation}
with,
\begin{align}
&H_{\mp} =\mathcal{L} +\epsilon \mp i\gamma \frac{\partial }{\partial x}-2\eta |\psi_{\pm 1}|^2, \notag \\
&H_0=\mathcal{L}-2\eta |\psi_{0}|^2, \notag\\
& \mathcal{L}= -\frac{1}{2}\frac{\partial^2}{\partial x^2} -\frac{V}{2}\cos(2x) +(1+\eta)n. \notag
\end{align}
In above GP equations, the spinor wave functions are $\psi(x,t)=[\psi_{-1}(x,t), \psi_0(x,t), \psi_1(x,t)]^T$, which describe the probability amplitudes of corresponding three hyperfine states $|-1\rangle, |0\rangle, |1\rangle$. The total density $n$ is defined as $n= |\psi_{-1}|^2 + | \psi_{0} |^2+ | \psi_{1} |^2$.
For the convenience of numerical calculations, the above GP equations are dimensionless. The units of length, energy, and time are $1/k_\mathrm{lat}$,  $2E_\mathrm{lat}=\hbar^2 k_\mathrm{lat}^2/m=2\pi\hbar \times 0.97\text{kHz}$,  and $\hbar/(2E_\mathrm{lat})$  respectively. The wave functions are also dimensionless with the unit being $\sqrt{ E_\mathrm{lat}/(\hbar \omega_\perp c_0) }$, here $\omega_\perp$ is the harmonic trap frequency along the transverse direction and $c_0=(a_0+2a_2)/3$ with $a_0$ and $a_2$ being s-wave scattering lengths in the total spin 0 and 2 channels~\cite{Ohmi1998,Ho1998}. With these units and experimental parameters of lasers ($\lambda_\text{Ram} =790 \text{nm}$ and $\lambda_\mathrm{lat}=1540\text{nm}$)~\cite{Campbell2016,Hamner2015}, the dimensionless quantities, the spin-orbit coupling strength $\gamma$ and the effective quadratic Zeeman shift $\epsilon$, become
\begin{align}
& \gamma=2 \frac{k_\mathrm{Ram}}{k_\mathrm{lat}}=2\sqrt{2}\frac{\lambda_\mathrm{lat}}{\lambda_\mathrm{Ram}} =5.52, \notag \\
&\epsilon= 2 \frac{k_\mathrm{Ram}^2}{k_\mathrm{lat}^2} + \frac{\delta}{2E_\mathrm{lat}} =15.2+\frac{\delta}{2E_\mathrm{lat}},
\label{Parameter}
\end{align}
and $V=\bar{V}/(2E_\mathrm{lat})$,  $\Omega=\bar{\Omega}/ (2E_\mathrm{lat})$, and $\eta=c_2/c_0$ with $c_2=(a_2-a_0)/3$. For $^{87} \text{Rb}$ atoms,  $a_0=101.8a_B$ and $a_2=100.4a_B$ with $a_B$ being the Bohr radius~\cite{Kempen2002}, therefore, 
\begin{equation}
\eta=\frac{c_2}{c_0}=\frac{a_2-a_0}{a_0+2a_2}=-0.005,
\end{equation}
which represents a ferromagnetic spin-spin interaction. We characterize the spin-orbit-coupled spinor gap soliton by its atom number 
\begin{equation}
N=N_0\int dx\left[  |\psi_{-1}|^2 + | \psi_{0} |^2+ | \psi_{1} |^2  \right],
\label{number}
\end{equation}
which is measured with $N_0=\hbar k_\mathrm{lat} /(  2m\omega_\perp c_0 )$. A typical experimental transverse trap frequency $\omega_\perp/(2\pi)= 300\text{Hz}$ leads to $N_0=105$.

The ground states of a spin-1 spinor BEC are polar states when $\eta>0$ and ferromagnetic states  when $\eta<0$~\cite{Ho1998}. Meanwhile, the ground states have an infinite spin degeneracy; all spinors associated with spin rotations are degenerate~\cite{Ho1998}.
The spin-1 spinor BEC can support stable bright solitons~\cite{Ieda2004,Luli2005,Szankowski2010}, which may have the spin degeneracy~\cite{Luli2005}. This may be understood from the single-mode approximation~\cite{Yi2002}. Three components share a same spatially localized profile which is decoupled from the spin degrees of freedom. Therefore, the spin rotation invariance of the spinor interactions is not affected under the single-mode approximation.  In the presence of optical lattices, spinor bright solitons exist inside the linear energy gaps, converting to gap solitons. Ref.~\cite{Kivshar2007} has found a family of spin-1 spinor gap solitons which have the same spatial profile in three components and have the spin degeneracy. Different from the ground states, the stable polar-like and ferromagnetic-like spinor gap solitons can exist no matter the sign of $\eta$. Moreover, Ref.~\cite{Kivshar2007} has found the existence of stable gap solitons that do not satisfy the single-mode approximation. 
 
  The coupling between the spin and orbit degrees of freedom leads to that the single-mode approximation generally cannot apply to spin-orbit-coupled spinor BECs. The ground states of a spin-1 spin-orbit-coupled spinor BEC have intriguing phases including stripe, plane-wave and zero-momentum states~\cite{Lan2014,Natu2015,Yu2016,Sun2016,Martone2016}.  Spin textures and spin dynamics~\cite{Mardonov2015} of spin-orbit-coupled spinor bright solitons become important due to the lacking of the spin degeneracy~\cite{Liu2014,Mardonov2018,Ma2019,Adhikari2019,Sun2020,Chengang2020, Adhikari2021,Meng2022}.

Here, we study spin-1 spin-orbit-coupled spinor gap solitons by numerically solving the Eq.~(\ref{GP}). In the experiment~\cite{Campbell2016}, the Raman coupling $\Omega$ and the two-photon detuning $\delta$ are tunable parameters. In our study, we keep  $\Omega$ as a free parameter and choose two typical values for the detuning $\delta=-30.4E_\mathrm{lat}$ and  $\delta=0$.  This leads to that the effective quadratic Zeeman term of the GP equations in Eq.~(\ref{GP}) and Eq.~(\ref{Parameter}) becomes  $\epsilon=0$ and $\epsilon=15.2$ correspondingly.  Due to the locations of gap solitons, it is important to identify linear energy gaps firstly.

\begin{figure*}[t]
	\includegraphics[width=7in]{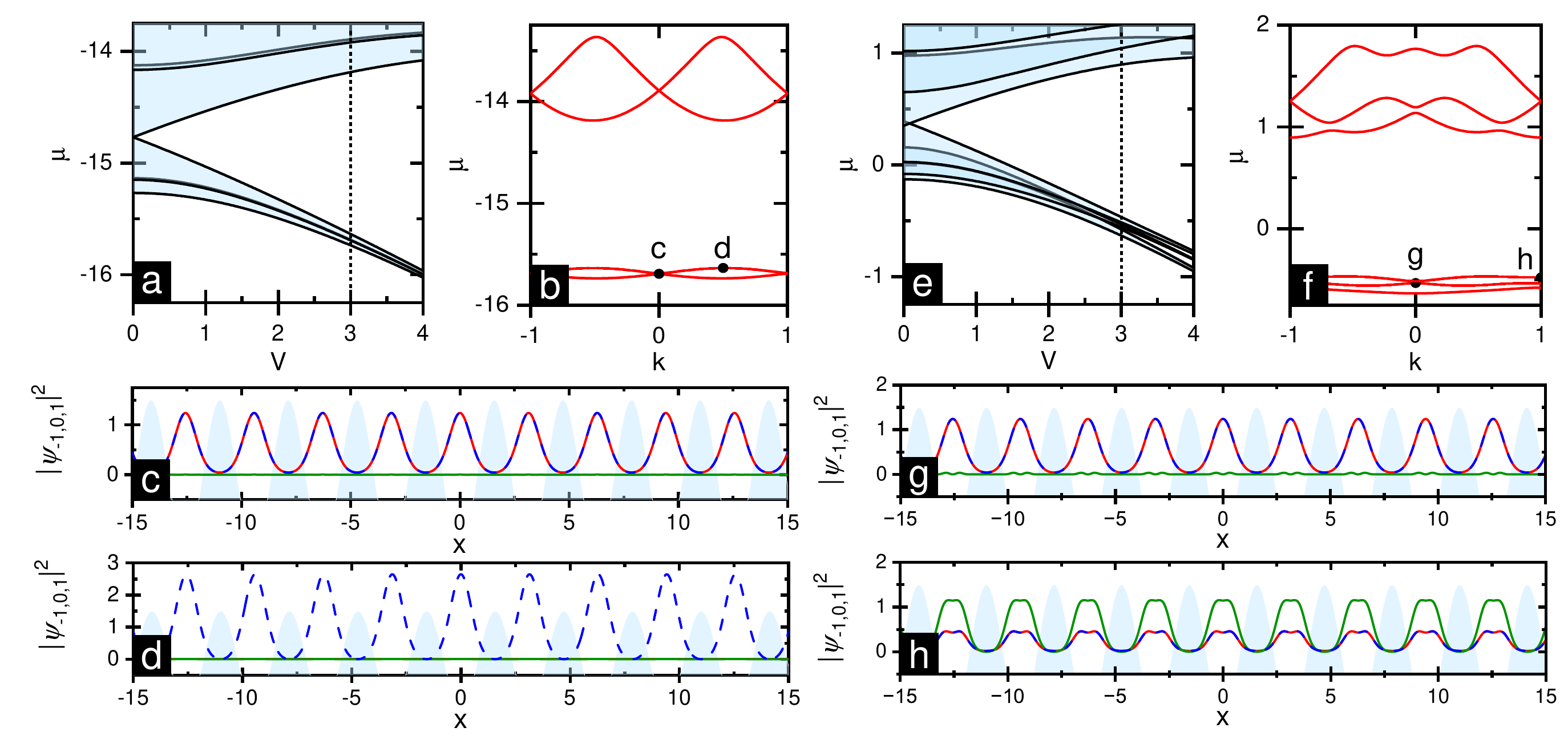}
	\caption{  Linear Bloch spectrum of the spin-1 spin-orbit-coupled spinor optical lattice and corresponding Bloch waves. The dimensionless parameters are $\gamma =5.52$ and $ \Omega=1$.  In (a)-(d), the effective quadratic Zeeman shift is $\epsilon=0$.   (a): The Bloch spectrum as a function of the optical lattice depth $V$. The shadow areas indicate Bloch energy bands and the white areas are energy gaps.  The solid lines are the maximum and minimum of each Bloch band.  (b): The Bloch spectrum as a function of the quasimomentum $k$ for a fixed depth $V=3$ [indicated by the vertical dashed line in (a)]. (c) and (d): The density distributions of the Bloch waves labeled by the solid circles in (b). The shadow areas indicate the optical lattice regimes with $-V\cos(2x)/2>0$. ($|\psi_{-1}|^2, |\psi_{0}|^2,|\psi_{1}|^2$) are labeled by the red, olive and blue lines respectively. In (c), $|\psi_{-1}|^2 =|\psi_{1}|^2$, and $|\psi_{0}|^2=0$. In (d), $|\psi_{-1}|^2=|\psi_0|^2=0$.  (e)-(h) show same quantities as in (a)-(d), but the effective quadratic Zeeman shift is $\epsilon=15.2$.  }
	\label{Fig2}
\end{figure*}

\section{Linear spectrum}
\label{Linear}

The optical lattice is periodic, so the single-particle Hamiltonian $H_\mathrm{sin}$ in Eq.~(\ref{singleparticle}) possesses Bloch band gap spectrum. The corresponding Bloch waves are defined as $\psi(x,t)=\exp(-i\mu t +ikx)\phi(x)$. Here $\phi(x)=[\phi_{-1}(x),\phi_{0}(x),\phi_{1}(x)]^T$ are periodic functions having the same period as the optical lattice, $k$ is the quasimomentum, and $\mu$ is the chemical potential. The linear Bloch spectrum $\mu(k)$ can be calculated by a plane-wave expansion of $\phi(x)$.

Two typical linear Bloch spectra for the effective quadratic Zeeman shift $\epsilon=0$ and  $\epsilon=15.2$ are shown in Fig.~\ref{Fig2}.  For the case of $\epsilon=0$ in Fig.~\ref{Fig2}(a), the lowest two bands mix together and there is an energy gap between the second and third bands. The size of this gap increases as the increase of the optical lattice depth. The lowest four bands of the linear spectrum at $V=3$ are shown in Fig.~\ref{Fig2}(b) as a function of the quasimomentum. Clearly the gap between the second and third bands is big enough in comparison with the band widths of the lowest two bands. It may provide an accommodation to support spinor gap solitons. The density distributions of the Bloch waves at $k=0$ and at the maximum of the second band [labeled by the solid circles in  Fig.~\ref{Fig2}(b)] are demonstrated in Figs.~\ref{Fig2}(c) and \ref{Fig2}(d) respectively. It is interesting to see that these Bloch waves have specific spin populations, i.e.,  $|\psi_{-1}|^2 =|\psi_{1}|^2$ and $|\psi_{0}|^2=0$ in  Fig.~\ref{Fig2}(c),  and $|\psi_{-1}|^2=|\psi_0|^2=0$ in Fig.~\ref{Fig2}(d). 

We depict the linear spectrum for a large effective quadratic Zeeman shift $\epsilon=15.2$ in Figs.~\ref{Fig2}(e)-(h). As shown in Fig.~\ref{Fig2}(e), the lowest three bands mix together and there is no energy gap between them when the depth $V$ is small. For a sufficient large $V$,  gaps are weakly opened between the lowest three bands.  There always is an energy gap opening between the third and fourth bands. Its size also increases as a function of $V$. Therefore, for the existence of spinor gap solitons in this gap, it is reasonable to choose a large $V$. Here we choose $V=3$. The linear spectrum as a function of the quasimomentum for $V=3$ is demonstrated in Fig.~\ref{Fig2}(f). As expected, the energy gap between the third and fourth bands is a proper accommodation for spinor gap solitons.  Furthermore, two typical Bloch waves are described in Figs.~\ref{Fig2}(g) and \ref{Fig2}(h). They are the Bloch waves at the Brillouin zone center and edge in the third band.  The outstanding feature of these Bloch waves is spin occupations; $|\psi_{0}|^2 \ll |\psi_{-1}|^2 =|\psi_{1}|^2$ in Fig.~\ref{Fig2}(g),  and  $|\psi_{0}|^2 > |\psi_{-1}|^2 =|\psi_{1}|^2$ in Fig.~\ref{Fig2}(h).

Once the energy gaps are identified, we numerically find  gap solitons located inside them. In the following, we shall focus on the energy gaps shown in Figs.~\ref{Fig2}(b) and \ref{Fig2}(f). The profiles of linear Bloch waves shown in Fig.~\ref{Fig2} are instructive for the structures of gap solitons since they may bifurcate from these linear waves.

\begin{figure}[!t]
	\includegraphics[width=3.45in]{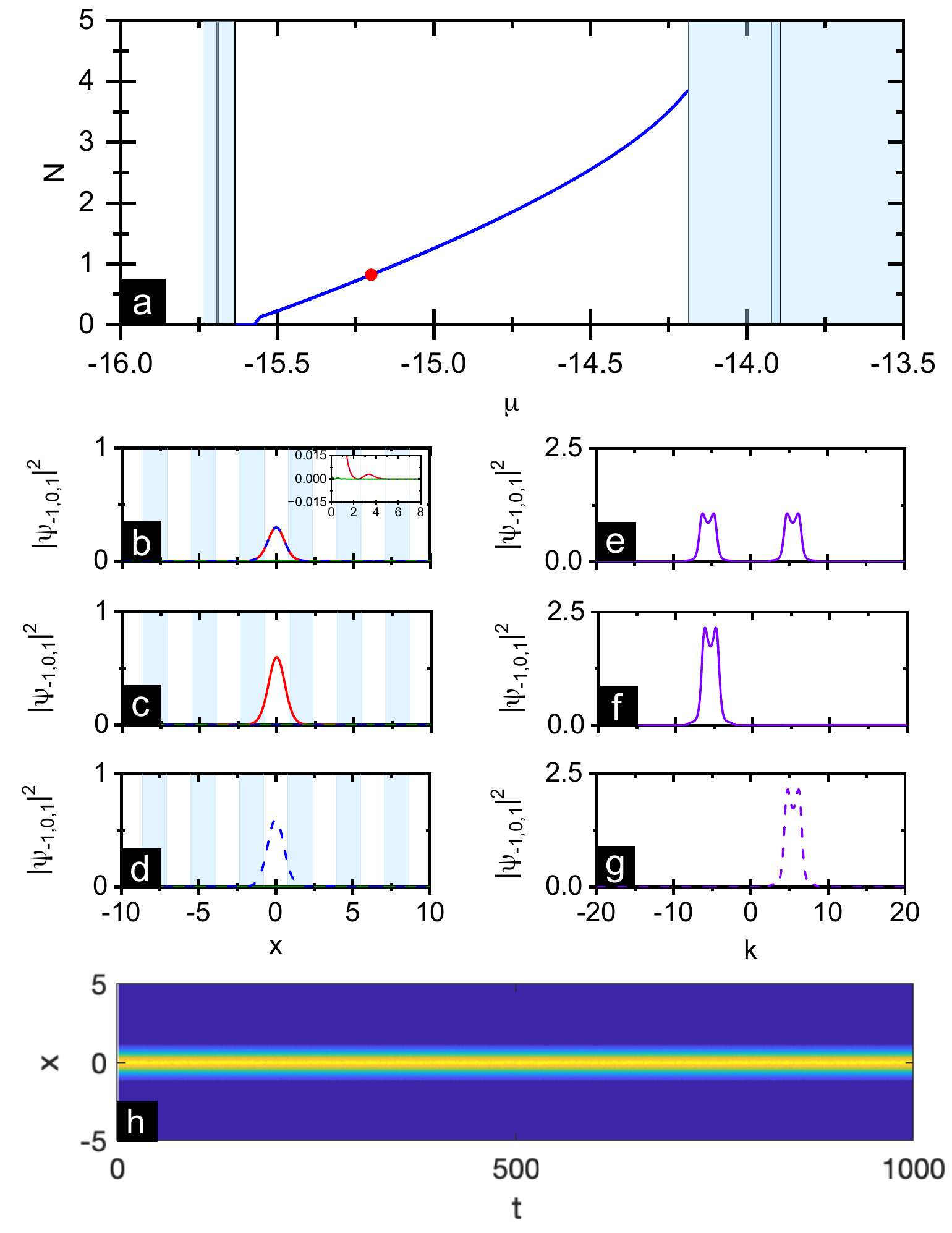}
	\caption{Two types of spin-1 spin-orbit-coupled spinor gap solitons in the case of the zero effective quadratic Zeeman shift $\epsilon=0$. The other parameters are the optical lattice depth $V=3$ and the Raman coupling $\Omega=1$. (a) The existence of spinor gap solitons is reflected by the dependence of the solitonic atom number $N$ on the chemical potential $\mu$. Two types are degenerate in the plane of ($N,\mu$). The shadow areas correspond to linear bands. The density profiles of the labeled point are shown in (b)-(d). (b) A density profile of the first type, $|\psi_{-1}|^2=|\psi_1|^2\ne 0$ and $|\psi_0|^2=0$, which is polar-like.  The inset zooms in the density oscillating tail. (c) Density profile of the second type, 
	$|\psi_{-1}|^2\ne 0$ (red line) and $|\psi_{0}|^2=|\psi_{1}|^2=0$.	 (d)  Density profile of the second type, 
	$|\psi_{-1}|^2=|\psi_{0}|^2=0$ and $|\psi_{1}|^2\ne 0$ (Blue dashed line). In (b)-(d), the red, olive, and blue-dashed lines correspond to $|\psi_{-1}|^2$, $|\psi_{0}|^2$ and $|\psi_{1}|^2$, respectively.  The shadow areas indicate the regimes for the lattices $-V\cos(2x)/2>0$. (e)-(g) The momentum-space distributions corresponding to (b)-(d). The $|-1\rangle$ ($|1\rangle$) component has a momentum peak centered at $k=-\gamma$ ($\gamma$). (h) The stable time evolution of a gap soliton. The initial state is $\psi(1+0.1\mathrm{Rand})$ with $\psi$ being the first-typed gap soliton labeled by the dot in (a) and $10\%$ Gaussian distributed  random noise is considered. }
	\label{Fig3}
\end{figure}

\section{Spin-orbit-coupled spinor gap soliton with the zero effective quadratic Zeeman shift}
\label{NegativeZeeman}

Spin-orbit-coupled spinor gap solitons are stationary solutions of the Eq.~(\ref{GP}).  $\psi(x,t)=\exp(-i\mu t)\psi(x)$ with the chemical potential $\mu$. Then, $\psi(x)$ satisfy the stationary GP equations. We use the Newton relaxation method to solve the spatially discretized stationary GP equations. The chemical potential is fixed to be the values inside the linear energy gaps during the calculations. We first consider the case of the zero effective quadratic Zeeman shift $\epsilon=0$. The solitons will locate inside the linear energy gap demonstrated in Fig.~\ref{Fig2}(b).

The results of  gap solitons are shown in Fig.~\ref{Fig3}. We find that the fundamental solitons have two different types. The dependence of solitonic atom number $N$ [defined in Eq.~(\ref{number})] on the chemical potential $\mu$ is shown in Fig.~\ref{Fig3}(a).  Two types are completely degenerate in the ($N,\mu$) plane. From this figure, we can know their existence; they exist inside the linear energy gap except a very small regime close to the second band where the wave functions go to zero.  Without the spin-orbit coupling, the wave functions of spinor gap solitons can be scaled into real numbers~\cite{Kivshar2007}. However, the presence of the spin-orbit coupling makes the scaling impossible, and the wave functions are in general complex numbers. The first type is $|\psi_{-1}|^2=|\psi_1|^2\ne 0$ and $|\psi_0|^2=0$, whose profile is demonstrated in Fig.~\ref{Fig3}(b). The second type has two configurations: $|\psi_{-1}|^2\ne 0$ and  $|\psi_0|^2=|\psi_1|^2=0$  as shown in Fig.~\ref{Fig3}(c), and $|\psi_{-1}|^2=|\psi_0|^2=0$ and $|\psi_{1}|^2\ne 0$ as shown in Fig.~\ref{Fig3}(d).
Since the degeneracy of these two types, the density amplitude of the second type is double that of the first. This is because that the second type has only one-component occupation, while the first type includes two components. No matter which type the soliton is, the common feature is the zero population of the $|0\rangle$ component,  $|\psi_0|^2=0$.

In the following, we provide a physical insight for these numerical solutions of  Eq.~(\ref{GP}). The effective quadratic Zeeman shift consists of two parts: one is due to the spin-dependent momentum displacement $ 2\hbar^2  k_\mathrm{Ram}^2/m$  and the other is the two-photon detuning $\delta$. $\epsilon=0$ means $\delta =-30.4E_\mathrm{lat}$, which is a large negative value. The momentum displacement does not affect the energy of system. The energy functional of the $\delta$ term is, $E_\text{qua}=\delta ( |\psi_{-1}|^2+ |\psi_{1}|^2 ) $. Since $\delta\ll 0$, minimizing this energy functional requires $\psi_0=0$. Physically, when the two-photon detuning $\delta$ is a large negative value we can adiabatically eliminate the $|0\rangle$ state.  Under the condition of $\psi_0=0$ and $\epsilon=0$, the  stationary GP equations in Eq.~(\ref{GP}) are simplified as,
\begin{align}
	&\mu \psi_{-1}= \mathcal{L'}  \psi_{-1}   - i\gamma \frac{\partial\psi_{-1}}{\partial x} +\eta (|\psi_{-1}|^2-|\psi_1|^2) \psi_{ -1}, \notag \\
	& \mu \psi_{1}= \mathcal{L'}  \psi_{1}   +  i\gamma \frac{\partial\psi_{1}}{\partial x} -\eta (|\psi_{-1}|^2-|\psi_1|^2) \psi_{ 1},
\end{align}
with
\begin{equation}
\mathcal{L'}=-\frac{1}{2}\frac{\partial^2}{\partial x^2}-\frac{V}{2}\cos(2x)+(|\psi_{-1}|^2+|\psi_1|^2) . \notag
\end{equation}
Note that the Raman coupling $\Omega$ disappears in the above reduced equations due to $\psi_0=0$. In the absence of $\Omega$, the spin-orbit coupling can be gauged out from the equations. After applying a unitary transformation $\psi_{-1}=\exp(-i\gamma x )\phi_{-1}$ and $\psi_{1}=\exp(i\gamma x )\phi_{1}$, the above equations become,
\begin{align}
	&\mu \phi_{-1}= \mathcal{L'}  \phi_{-1}  +\eta (|\phi_{-1}|^2-|\phi_1|^2) \phi_{ -1}, \notag \\
	& \mu \phi_{1}= \mathcal{L'}  \phi_{1}  -\eta (|\phi_{-1}|^2-|\phi_1|^2) \phi_{ 1}.
	\label{effective}
\end{align}
An irrelevant constant $\gamma^2$ has been dropped off.  The energy functional of the nonlinear terms in the above equations is \begin{equation}
E_\text{non}=\frac{1}{2}(1+\eta)(|\phi_{-1}|^4+|\phi_1|^4)+(1-\eta)|\phi_{-1}|^2|\phi_1|^2. \notag
\end{equation}
Since $\eta=-0.005<0$, the above interactions are immiscible. In the absence of the optical lattice,  to minimize the immiscible interactions, the ground state has two configurations: $\phi_{-1}\ne 0$, $\phi_{1}= 0$ and $\phi_{-1}=0$, $\phi_{1}\ne 0$. The ground state chooses the one of two configurations spontaneously. Therefore, considering the zero $|0\rangle$ component $\psi_0=0$, the ground state of the spin-orbit coupled spinor BEC without the optical lattice is ferromagnetic and is a plane wave as  $\psi_{-1}=\exp(-i\gamma x )\phi_{-1}$, $\psi_{1}=\exp(i\gamma x )\phi_{1}$ for the situation of the zero effective quadratic Zeeman shift. This result is consistent with that in Ref.~\cite{Campbell2016}.  

In the presence of the optical lattice, gap soliton solutions of the Eq.~(\ref{effective}) can be numerically found as $\phi_{-1}^{(gs)}(x)$ and  $\phi_{1}^{(gs)}(x)$.  Since all quantities in Eq.~(\ref{effective}) are real-valued, the wave functions of gap solitons $\phi_{-1}^{(gs)}$ and  $\phi_{1}^{(gs)}$ can be scaled into real numbers. On the other hand, the conservation of the total density $|\phi_{-1}|^2+|\phi_{1}|^2  $ and the relative density $|\phi_{-1}|^2-|\phi_{1}|^2 $ in the time-dependent version of Eq.~(\ref{effective}) brings two free degrees of freedoms into the gap soliton solutions; they are the global phase $\alpha$ and relative phase $\theta$. Finally, going back to the basis of $\psi$, we get the solutions of gap solitons as
\begin{equation}
	\begin{pmatrix}
		\psi_{-1} \\ \psi_{0} \\ \psi_{1}
	\end{pmatrix}_{(gs)}= 	\begin{pmatrix} e^{-i\gamma x} \phi_{-1}^{(gs)} (x)\\ 0 \\ e^{i\gamma x}\phi_{1}^{(gs)}(x) e^{i\theta}	\end{pmatrix} e^{i\alpha}.
\label{solution}
\end{equation}

The numerical results demonstrated in Fig.~\ref{Fig3} belongs to the solutions in Eq.~(\ref{solution}). In order to show this, we plot the momentum-space density distributions of corresponding gap solitons in Figs.~\ref{Fig3}(e)-(g). The $|-1\rangle$ component of solitons always has a momentum peak locating at $k=-\gamma$, while it sits at $ k=\gamma $ in the $|1\rangle$ component. Results of momentum peaks are consistent with solutions in Eq.~(\ref{solution}). It is noticed that there is a small amplitude splitting in the each peak. The splitting originates from the existence of very small oscillating tails in gap solitons which are shown in the zooming-in inset in Fig.~\ref{Fig3}(b).

The spin-orbit-coupled Hamiltonian in Eq.~(\ref{socH}) obeys a spin-dependent parity symmetry $\mathcal{O}$,
\begin{equation}
\mathcal{O}=-\mathcal{P} e^{i\pi F_x}= \mathcal{P}\begin{pmatrix}
	0&0&1 \\ 0&1 &0 \\ 1&0&0
\end{pmatrix}.
\end{equation}
Here, $\mathcal{P}$ is the parity operator, $\mathcal{P} x \mathcal{P}^{-1}=-x$, and $ e^{i\pi F_x}$ is the operator to rotate spins along the $F_x$ by an angle of $\pi$. The first type of spin-orbit-coupled spinor gap solitons satisfies this symmetry, which gives rise to $\psi_{-1}(x)=\pm \psi_1(-x)$.  This restriction in the end leads to $\phi_{-1}^{(gs)} (x) = \phi_{1}^{(gs)}(x)$ and $\theta$ is fixed to $0$ or $\pi$. While the global phase in the Eq.~(\ref{solution}) still cannot be fixed and $\alpha$ can be arbitrary real constants since the GP equations have the U(1) symmetry.   The second type wants the one of $\phi_{-1}^{(gs)}$ and $\phi_{1}^{(gs)}$  to disappear. Therefore, it does not obey the $\mathcal{O}$ symmetry.

 Finally, we calculate the spin average value for the gap solitons. It is defined as,
  \begin{equation}
 	\langle \mathbf{F} \rangle=  \int dx \psi^{T*}(x) \mathbf{F} \psi(x).
 \end{equation}
The first type satisfies $\langle \mathbf{F} \rangle=0$, which is reminiscent of the spinor polar ground state in the case of $\eta>0$~\cite{Ho1998}. However,  for spin-orbit-coupled spinor gap solitons, the first type is polar-like but can exist when $\eta<0$. While, the second type is ferromagnetic-like, $|\langle F_z \rangle|\ne 0$ and $\langle F_x \rangle=\langle F_y \rangle=0$.

The stability of gap solitons found in Fig.~\ref{Fig3}(a) is examined by the nonlinear evolution of the GP equations. The nonlinear evolution is implemented by using the initial states as $\psi(1+0.1\mathrm{Rand})$ with $\psi$ being gap solitons and $\mathrm{Rand}$ being the randomly distributed noise. The stable solutions shall evolve without changing density profiles. 	The results are that all gap soliton solutions in Fig.~\ref{Fig3}(a) are stable except for a very small regime that the chemical potential is close to the second band. In this regime, both types of solitons have a weak oscillation. In Fig.~\ref{Fig3}(h), we demonstrate a typically stable evolution of the first-typed soliton [which is labeled by the dot in Fig.~\ref{Fig3}(a)]. It shows that the corresponding soliton evolves stably up to $t=1000$ [which corresponds to 165ms considering the time unit] in the presence of noise.

\begin{figure}[t]
	\includegraphics[width=3.45in]{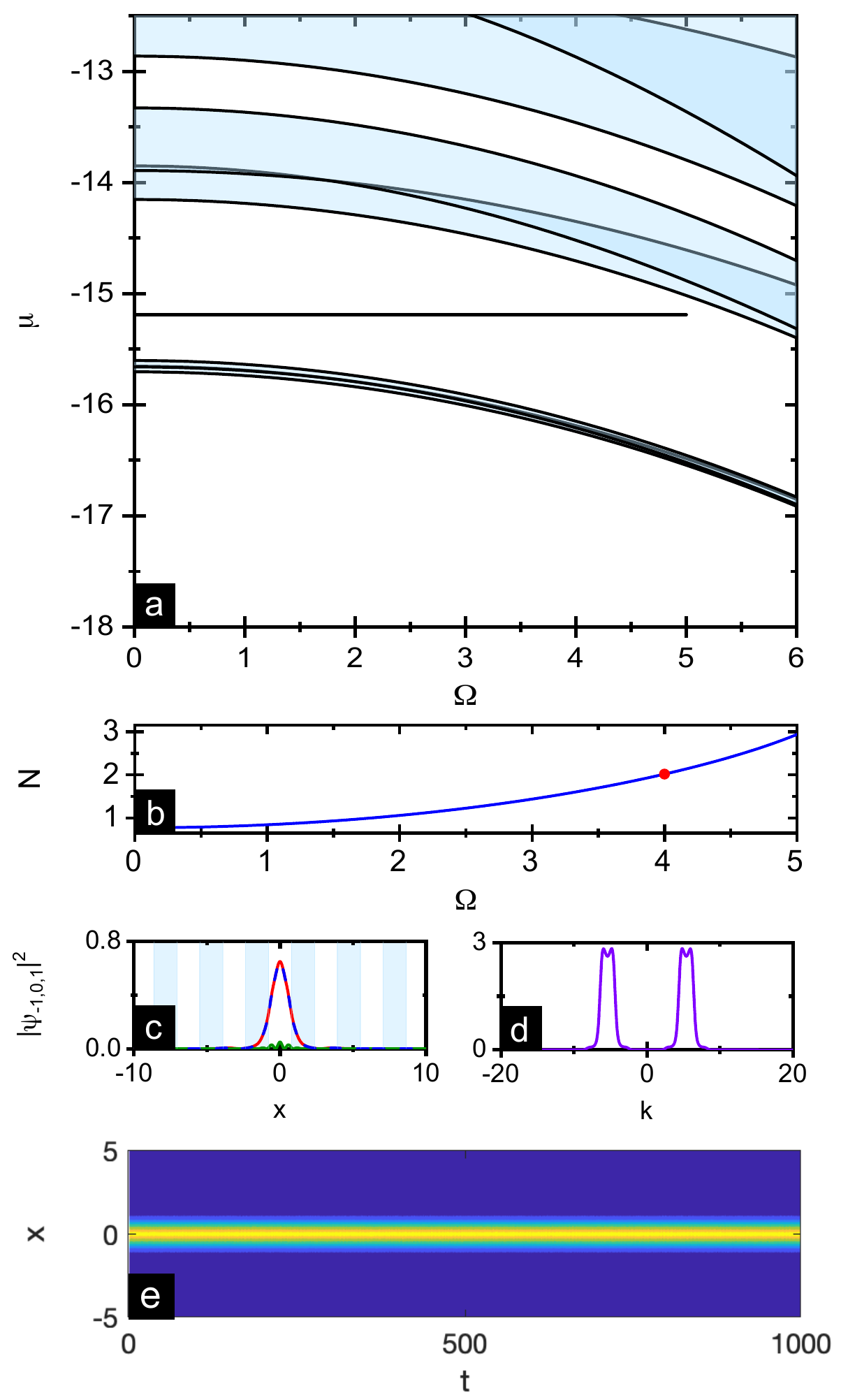}
	\caption{Two types of spin-1 spin-orbit-coupled spinor gap solitons with the zero effective quadratic effect $\epsilon=0$ can exist in a broad range of the Raman coupling $\Omega$. The optical lattice depth is $V=3$. (a) Linear Bloch spectrum as a function of  $\Omega$. The shadow areas represent energy bands.  (b) The dependence of the solitonic atom number on $\Omega$ for a fixed chemical potential [labeled by the horizontal line in (a)]. The two types are completely degenerate in the ($N,\Omega$) plane. The coordinate-space and momentum-space density distributions of the first-typed gap solitons [labeled by the dot in (b)] are demonstrated in (c) and (d) respectively.  The shadow areas in (c) represent $-V\cos(2x)/2>0$. (e) The stable time evolution of the first-typed gap soliton [labeled by the dot in (b)] with $10\%$ random noise.}
	\label{Fig4}
\end{figure}

These  polar-like and ferromagnetic-like types can exist for a broad range of the Raman coupling $\Omega$ as long as the linear energy gap keeps opening.  The evolution of linear spectrum as a function of $\Omega$ is demonstrated in Fig.~\ref{Fig4}(a). It shows that the energy gap between the second and third bands is widely opened as the change of $\Omega$. We fix the chemical potential and numerically search for gap solitons along the horizontal line shown in Figs.~\ref{Fig4}(a). The results are described in Fig.~\ref{Fig4}(b). Two types are still degenerate in the $(N, \Omega)$ plane. The solitonic atom number $N$ increases as the increase of $\Omega$. When $\Omega$ is small, the two-photon detuning $\delta$ completely dominates leading to the zero occupation in the $|0\rangle$ component. Since the Raman coupling works via the $|0\rangle$ component, without which, it has no effect on solitons. Therefore,  the number $N$ keeps as a constant for small $\Omega$. However, as it increases, the Raman coupling manages to flip spins to allow for the small occupation in the $|0\rangle$ component.  This causes to the increase of solitonic atom number with the increase of the Raman coupling as shown in Fig.~\ref{Fig4}(b). The typical coordinate-space and momentum-space density profiles of the first-typed spinor gap soliton at a large Raman coupling $\Omega=4$ are shown in Fig.~\ref{Fig4}(c) and \ref{Fig4}(d). This soliton looks like the previous result of $\Omega=1$ in  Fig.~\ref{Fig3}. However, the difference between them is that there is a small occupation in the $|0\rangle$ component [see the olive line in Fig.~\ref{Fig4}(c)].
The Raman coupling transfers a small amount of atoms to the $|0\rangle$ component from the $|-1\rangle$ and $|1\rangle$ components, and during the transfer momentum is conserved. Consequently, the $|0\rangle$ component becomes a superposition of two very small packets locating at $-\gamma$ and $\gamma$ in the momentum space. 

The stability of the whole branch in Fig.~\ref{Fig4} is also checked by the nonlinear evolution. All solitons are stable. We present a typically stable evolution of the first-typed solution [represented by the dot in Fig.~\ref{Fig4}(b)] in  Fig.~\ref{Fig4}(e). The soliton can exist stably for a very long time.

\section{Spin-orbit-coupled spinor gap solitons with a large effective quadratic Zeeman shift}
\label{ZeroZeeman}

In above, due to a very large negative $\delta$, the occupation of the $|0\rangle$ component is almost negligible, $|\psi_{0}|^2=0$. The first type of spinor gap solitons has $|\psi_{-1}|^2=|\psi_{1}|^2$ obeying the spin-dependent parity symmetry $\mathcal{O}$ .  The second type spontaneously chooses to occupy the one of $|\psi_{-1}|^2$ and $|\psi_{1}|^2$, therefore, is ferromagnetic-like. This type  breaks the spin-dependent parity symmetry. In the following, we consider  $\delta=0$, which leads to a large effective quadratic Zeeman shift $\epsilon=15.2$ in the GP equations. There is no constraint on the occupation of the $|0\rangle$ component. The gap solitons are searched inside the energy gap shown in Fig.~\ref{Fig2}(f).

\begin{figure}[t]
	\includegraphics[ width=3.45in]{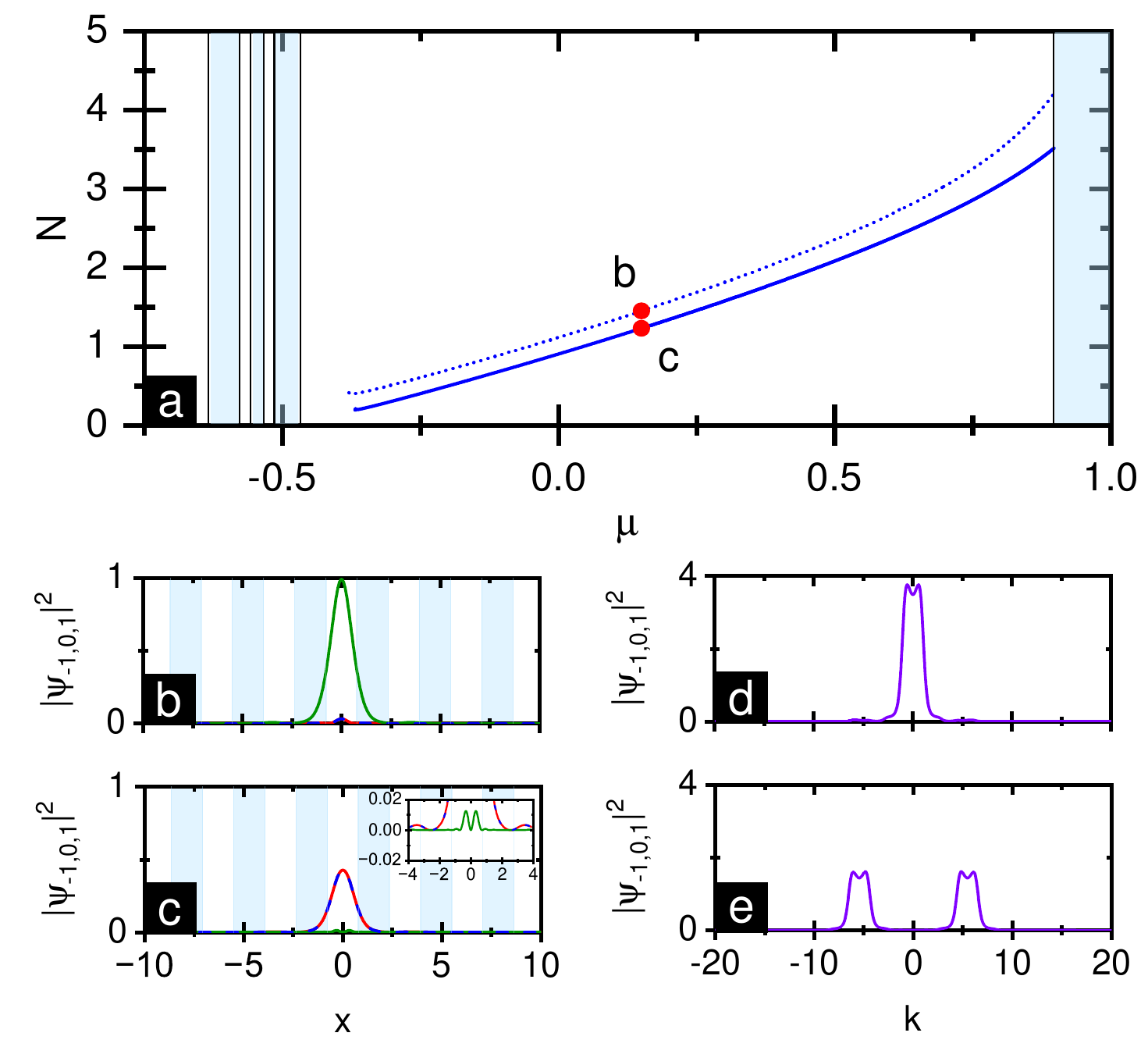}
	\caption{ Two different families of spin-1 spin-orbit-coupled spinor gap solitons with a large effective quadratic Zeeman shift $\epsilon=15.2$ [i.e., $\delta=0$]. Other parameters are $V=3$ and $\Omega=1$.
		(a) The existence of two families in the ($N,\mu$) plane. The first (second) family is represented by the dotted (solid) line. The wave functions of the labeled points shall be shown in (b) and (c). 
		The shadow areas correspond to linear bands. (b) A density profile of the first family in the coordinate space, $|\psi_{-1}|^2=|\psi_1|^2\ll |\psi_0|^2$. (c) A density profile of the second family in the coordinate space, $|\psi_{-1}|^2=|\psi_1|^2\gg |\psi_0|^2$. In (b) and (c), the red,  olive, and blue-dashed lines correspond to $|\psi_{-1}|^2$, $|\psi_{0}|^2$ and $|\psi_{1}|^2$, respectively.  The shadow areas indicate the regimes for the lattices $-V\cos(2x)/2>0$. In (c),  the inset zooms in the density distributions around $x=0$. 
		(d) and (e) The momentum-space-density distributions correspond to (b)-(c). }
	\label{Fig5}
\end{figure}

We find that there are two different families for the fundamental gap solitons. They are demonstrated in Fig.~\ref{Fig5}. The two families have different atom numbers in the ($N,\mu$) plane; the first family has a slightly larger  number [see the dotted line in Fig.~\ref{Fig5}(a)]. Both families exist in whole energy gap, excepting for a regime very close to the third linear band. The typical density profiles of the first family are shown in 
Figs.~\ref{Fig5}(b) and~\ref{Fig5}(d) in the coordinate and momentum space respectively. The first family features a dominating occupation of the $|0\rangle$ component, and the $|-1\rangle$ and $|1\rangle$ components have the same density with a very small population,  $|\psi_0|^2\gg |\psi_{-1}|^2= |\psi_1|^2$. The momentum of the $|0\rangle$ component lays at $k=0$. This can be seen from the momentum-density peak in Fig.~\ref{Fig5}(d) [noting that the small amplitude splitting still exists due to the tail of solitons]. The density profiles of the second family are described in Figs.~\ref{Fig5}(c) and~\ref{Fig5}(e). This family features $|\psi_{-1}|^2= |\psi_1|^2 \gg |\psi_0|^2$. The momentum of the $|-1\rangle$ component lays at $k=-\gamma$, and the $|1\rangle$ component is peaked around $k=\gamma$. It is very interesting to note that both families respect the spin-dependent parity symmetry $\mathcal{O}$.  The symmetry wants to wave functions to satisfy $\mathcal{O}\psi(x)=\pm \psi(x)$ with the eigenvalues being $\pm 1$, leading to
\begin{equation}
	\psi_{-1}(x)= \pm \psi_1(-x), \ \psi_0(x)= \pm \psi_0(-x).
\end{equation}
This result further gives rise to $\langle F_y \rangle=\langle F_z \rangle=0$. The first family belongs to the eigenstate with $+ 1$ eigenvalue, i.e., even spin-dependent parity,  this can be inferred from that there is no node in the density of the $|0\rangle$ component as shown by the olive line in Fig.~\ref{Fig5}(b). While the second family is odd spin-dependent parity with the eigenvalue $-1$, as there is a node at $x=0$ in the density of the $|0\rangle$ component [see the olive line in the inset in Fig.~\ref{Fig5}(c)].

\begin{figure}[b]
	\includegraphics[ width=3.4in]{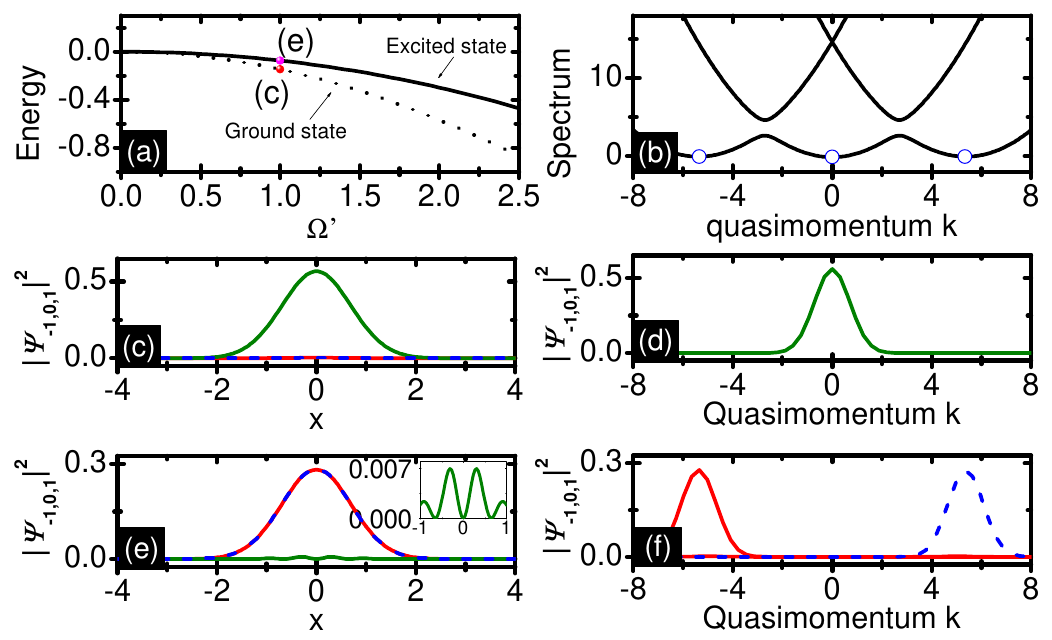}
	\caption{  (a) The ground state and first excited state of $H'_\text{sin}$ in Eq.~(\ref{Oscillator}). The wave functions of labeled points shall be shown in the following plots.  (b) The spectrum of $H_\text{soc}$ in the momentum space. Three local minima in the lowest band are labeled by circles. (c) and (d) The density distributions of a ground state in the coordinate and momentum space respectively. The olive line is $|0\rangle$ component,  and the red (blue-dashed)  line is the $|-1\rangle$ ($|1\rangle$) component. (e) and (f) The density distributions of a first excited state in the coordinate and momentum space. In (e), the inset shows the zooming-in of the density distribution of the $|0\rangle$ component around $x=0$.}
	\label{Fig6}
\end{figure}

The spin-dependent parity symmetry of the two families indicates their origin. We note that densities of these two fundamental families are mainly confined inside a unit cell of the optical lattice. To give a simple physical picture for these two families, we develop an approximate model to replace the optical lattice by a Harmonic trap. With the approximation, the single-particle Hamiltonian becomes,
\begin{equation}
H'_\text{sin}= H_\text{soc} + \frac{1}{2}m \omega^2 x^2,
\label{trap}
\end{equation}
where, $\omega$ is the Harmonic trap frequency. We use the Harmonic oscillator basis to diagonalize $H'_\text{sin}$ by introducing $x=\sqrt{\hbar /(2m\omega)} (a^\dagger+a)$ and $p_x=i\sqrt{\hbar \omega m/2}(a^\dagger-a)$, here $a$ and $a^\dagger$ are the annihilation and creation operators. Consequently, the single-particle Hamiltonian becomes,
\begin{align}
 \frac{H'_\text{sin}}{\hbar \omega}= 
  a^\dagger a+ i\frac{\gamma'}{\sqrt{2}} (a^\dagger -a )F_z +  \frac{\gamma'^2}{2} F_z^2 
 +\sqrt{2} \Omega' F_x.
 \label{Oscillator}
\end{align}
We have renormalized the energy by using the energy unit as $\hbar \omega$. The spin-orbit coupling coefficient is $\gamma'= 2k_\text{Ram}\sqrt{\hbar/(m\omega)}$ and $\Omega'=\bar{\Omega}/(\hbar \omega)$. Using the Harmonic trap $\omega/(2\pi)=1$kHz, we have $\gamma'=5.4$.  Diagonalizing the Hamiltonian in Eq.~(\ref{Oscillator}) gives rise to the energy and associated eigenstates.  The energies of ground state and the first excited state are shown in Fig.~\ref{Fig6}(a). When $\Omega'=0$, these two states are degenerate.  They decrease as the increase of $\Omega'$. The wave functions of the ground state and the first excited state are demonstrated in Figs.~\ref{Fig6}(c),(d) and~\ref{Fig6}(e),(f) respectively. It is well-known that the eigenstates of a Harmonic trapped system have the well-defined parity symmetry and neighboring two states have opposite signs of parity. In the presence of the spin-orbit coupling, the symmetry becomes the spin-dependent parity $\mathcal{O}$. The spin-dependent parity of the ground state is even and it is odd for the first excited state. This can be inferred from the density distributions of the $|0\rangle$ component in Figs.~\ref{Fig6}(c) and~\ref{Fig6}(e); for the ground state, there is no node in the density of the $|0\rangle$ component, while for the first excited state there is a node at $x=0$ in the $|0\rangle$ component [see the olive lines in Fig.~\ref{Fig6}(c) and in the inset in  Fig.~\ref{Fig6}(e)]. The ground state has only a population in the $|0\rangle$ component. While the first excited state has an equal occupation of the $|-1\rangle$ and $|1\rangle$ components and has a little occupation in the $|0\rangle$ component. 

\begin{figure}[t]
	\includegraphics[ width=3in]{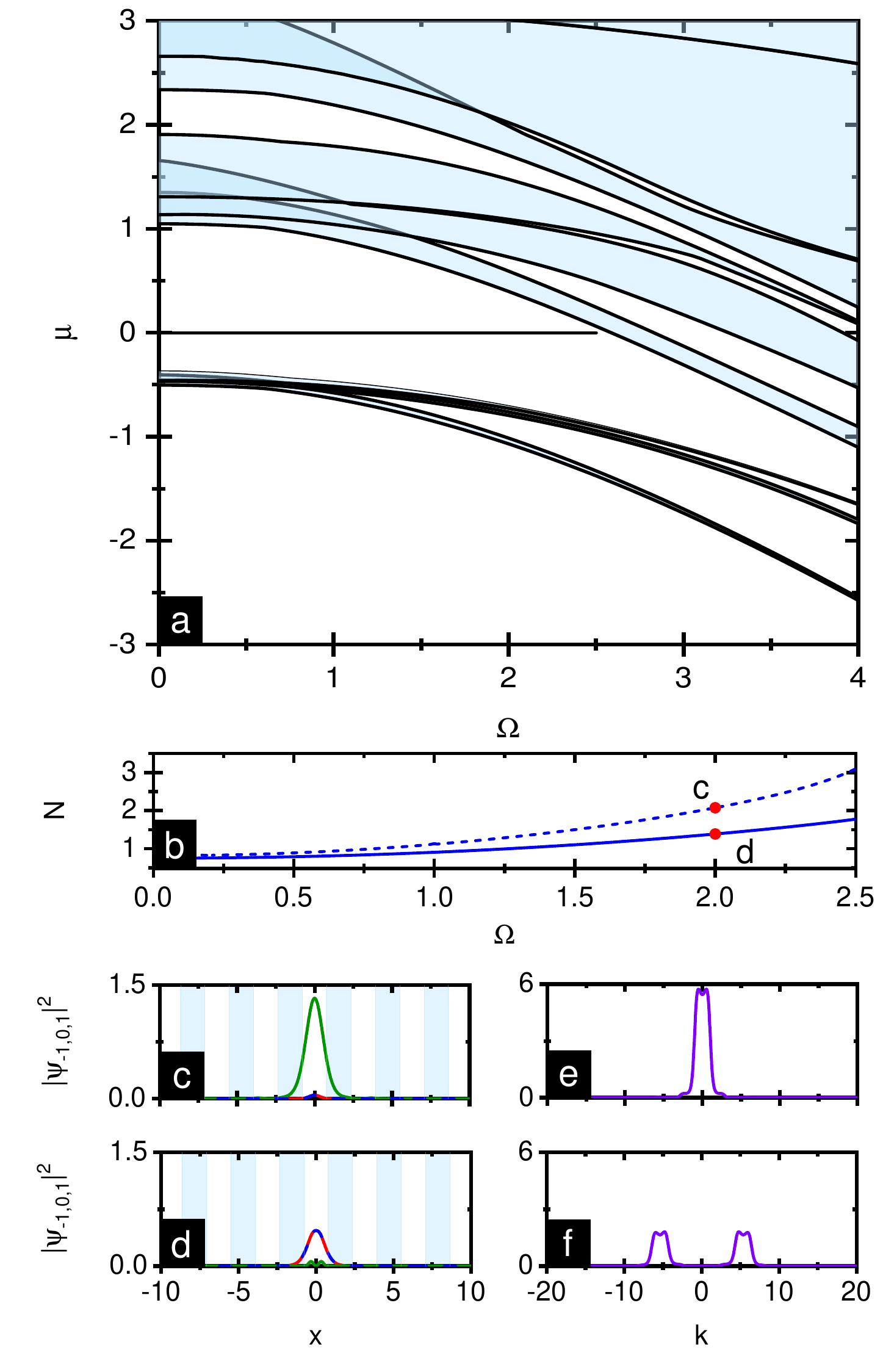}
	\caption{Two families of fundamental spin-orbit-coupled spinor gap solitons with a large effective quadratic effect $\epsilon=15.2$ can exist in a broad range of the Raman coupling $\Omega$. The optical lattice depth is $V=3$. (a) Linear Bloch spectrum as a function of  $\Omega$. The shadow areas represent energy bands. We shall search for gap solitons along the horizontal line. (b) The dependence of the solitonic atom number on $\Omega$ for a fixed chemical potential [labeled by the horizontal line in (a)]. The dotted (solid) line is for the first (second) family. The coordinate-space and momentum-space density distributions of two families [labeled by the dots in (b)] are demonstrated in (c), (e) and (d),(f) respectively.  The shadow areas in (c) and (d) represent $-V\cos(2x)/2>0$. }
	\label{Fig7}
\end{figure}

Without the Harmonic trap, the Hamiltonian $H_\text{soc}$ can be directly diagonalized in the momentum space. The spectrum of $H_\text{soc}$ has three energy bands, as shown in  Fig.~\ref{Fig6}(b). The lowest band features a three-parabolic-like structure with three local minima which are labeled by circles in the figure. The right and left minima are degenerate and their energy  is a bit higher than the middle one. The ground state of $H'_\text{sin}$ comes from the occupation of the middle minimum. Since the middle parabola is mainly generated by the $|0\rangle$ component,  the ground state possesses the dominating population in the $|0\rangle$ component and its momentum is peaked at $k=0$, which is demonstrated in Fig.~\ref{Fig6}(d). The first excited state of $H'_\text{sin}$ originates from an equal superposition of the right and left minima. In order to take the odd spin-dependent parity symmetry, the superposition should be out of phase. As the right (left) parabola is generated by the $|-1\rangle$ ($|1\rangle$) component, the first excited state becomes  $\propto \exp(-ik_{\min}x)(\psi_{-1},0,0)^T - \exp(ik_{\min}x)(0,0,\psi_1)^T$, where $ k_{\min}= \gamma'$ is the location of the left minimum. Therefore, in the momentum space, the first excited state features two momentum peaks locating at $\pm \gamma'$, which are demonstrated in Fig.~\ref{Fig6}(f).

The first (second) family of spin-orbit-coupled spinor gap solitons is strongly reminiscent of  the ground state (first excited state) of the Harmonic trapped system.  For a fixed atom number, the chemical potential of the first family should be smaller than the second family. This causes that the first family lays above the second one in the $(N,\mu)$ plane in Fig.~\ref{Fig5}(a). All features of two families, such as the spin-dependent parity symmetry, spin composition and momentum distribution, are the same as those of the ground state and first excited state. Therefore, the features are single particle effect and are induced solely by the spin-orbit coupling. While the spinor interactions guarantee the existence of gap solitons.

By examining the stability of these two families, we find that the whole branch of the second family in Fig.~\ref{Fig5}(a) is dynamically stable. While the first family is also stable, but except for a very small regime close to the third band where they are weakly unstable. 

The two families with the opposite spin-dependent parity can exist and are stable in a wide range of the Raman coupling $\Omega$. The linear Bloch spectrum as a function of $\Omega$ is shown in Fig.~\ref{Fig7}(a). It demonstrates that the size of the energy gap between the third and fourth bands decreases as the increasing of $\Omega$. Therefore, the existence of gap solitons in this gap is not preferred in the regime of large $\Omega$. We find their existence in the regime of $0 \le \Omega \le 2.5 $ for a fixed chemical potential as $\mu=0$. The dependence of the solitonic atom number $N$ on $\Omega$ is demonstrated in Fig.~\ref{Fig7}(b). The first family is always higher than the second one. As $\Omega$ increases, the difference of the atom number between two families becomes large. This can be understood from the approximate model of the Harmonic trap. The energy of the ground state decreases more dramatically than the first excited state as the increase of $\Omega$, which is described in Fig.~\ref{Fig6}(a). For the gap solitons in Fig.~\ref{Fig7}(b), the chemical potential of two families is fixed to be equal. The atom number must be adjusted to generate the same effect as the energy in the Harmonic trapped case. Typical coordinate-space and momentum-space densities at $\Omega=2$ are shown in Figs.~\ref{Fig7}(c),(e) for the first family and in Figs.~\ref{Fig7}(d),(f) for the second one. All features of densities are the same as these shown in Figs.~\ref{Fig5}(b)-(e).

 \section{conclusion}
 \label{Conclusion}

Stimulated by the experimental realization of the spin-1  spin-orbit-coupled BEC in Ref.~\cite{Campbell2016}, we study the spin-orbit-coupled spinor gap solitons in this system. They can exist in the wide energy gaps of corresponding  linear system. The spinor interactions possess the spin rotation symmetry. While the spin-orbit coupling breaks this symmetry by respecting a particular spin-dependent parity symmetry which is a specific spin rotation accompanying with an orbital parity. In the case of  a large effective quadratic Zeeman shift $\epsilon$, we uncover two different families of gap solitons having opposite spin-dependent parity. By developing a spin-orbit-coupled model with a Harmonic trap,  these two families of gap solitons can be considered as the ground state and first excited state of the Harmonic trapped system. The spin-orbit coupling brings intriguing features, including spin compositions and momentum distributions,  to the two families. For the zero effective quadratic Zeeman shift, the two-photon detuning $\delta$ dominates, which prevents the population in the $|0\rangle$ component. In this case, we reveal that spin-orbit-coupled spinor gap solitons have two types; one type obeys the spin-dependent parity symmetry and is polar-like, and the other type spontaneously breaks the symmetry and is ferromagnetic-like. All gap solitons in our study are fundamental modes whose main density peaks confines inside a unit cell of the optical lattice.  Using the fundamental modes as building blocks, higher-order modes with rich spatial structures are expected to be constructed.

 In our parameter regimes, there always exists a kind of  gap solitons. They occupy the $|-1\rangle$ and $|1\rangle$ components and have a negligible population in the $|0\rangle$ component. Using the idea of synthetic dimension, the $|-1\rangle$ and $|1\rangle$  components are the two edges of three-leg ladders. Therefore, this kind of solitons belongs to edge states.  However, different from usual edge states which are extended waves along the edges, due to solitary property, these states are also spatially localized along the edges. Physically, the spin-orbit coupling induces a nonzero magnetic flux to the three-leg ladders. The magnetic flux guarantees the existence of the edge states which can have a dispersion along the edges. While, the nonlinearities can be used to balance the dispersion to generate solitons along the edges. It is of interest that for the zero effective quadratic Zeeman shift, such ``edge state solitons" can spontaneously choose to occupy only the one of two edges.  Our study fully relies on the experimental system and all parameters are within the experimental accessibility. These make it possible to observe  ``edge state solitons" experimentally in the future. 

\section{Acknowledges}
This work was supported by National Natural Science Foundation of China with Grants No.11974235 and
11774219.

\end{document}